\documentclass[prd,nofootinbib,showpacs,preprint]{revtex4}
\usepackage{graphicx}
\usepackage{epsfig}
\usepackage{amsmath}
\usepackage{amsfonts}
\usepackage{amssymb}
\usepackage{url}
\usepackage{hyperref}
\usepackage{subfigure}
\newcommand{\beqa}{\begin{eqnarray}} 
\newcommand{\eeqa}{\end{eqnarray}} 
\newcommand{\beq}{\begin{equation}} 
\newcommand{\eeq}{\end{equation}}

\newcommand{\bmt}{\begin{pmatrix}}
\newcommand{\emt}{\end{pmatrix}}
\newcommand{\be}{\begin{equation}}
\newcommand{\ee}{\end{equation}}
\newcommand{\bea}{\begin{eqnarray}}
\newcommand{\eea}{\end{eqnarray}}

\begin{document}
\title{Dipolar dark matter in light of 3.5 keV X-ray Line, Neutrino mass and LUX data}
\author{Sudhanwa Patra$^{1}$, Nirakar Sahoo$^{2}$ and Narendra Sahu$^{2}$}
\email{sudha.astro@gmail.com, ph13p1005@iith.ac.in, nsahu@iith.ac.in}
\affiliation{$^{1}$\,Centre of Excellence in Theoretical and Mathematical Sciences, Siksha \textquoteleft O\textquoteright 
             Anusandhan University, Bhubaneswar, India\\
             $^{2}$\,Department of Physics, Indian Institute of Technology, 
             Hyderabad, Yeddumailaram, 502205, Telengana India}
\begin{abstract}
A simple extension of the standard model (SM) providing transient magnetic moments to right-handed neutrinos 
is presented. In this model, the decay of next-to-lightest right-handed heavy neutrino to the lightest one and a photon 
($N_2\to N_1 + \gamma$) can explain the $\sim$ $3.5$ keV X-ray line signal observed by XMM-Newton 
X-ray observatory. Beside the SM particles and heavy right-handed Majorana neutrinos, the model contains a singly charged 
scalar ($H^\pm$) and an extra Higgs doublet ($\Sigma$). Within this minimal set of extra fields the sub-eV masses of 
left-handed neutrinos are also explained. Moreover, we show that the spin-independent DM-nucleon cross-section is 
compatible with latest LUX data.
\end{abstract}
\maketitle
\section{Introduction}\label{sec:intro}
The existence of dark matter, which constitutes around 26.8\% of the total energy budget of the 
Universe~\cite{PLANCK}, is currently supported by the galaxy rotation curve, gravitational lensing and 
large scale structure of the Universe~\cite{DM_review}. These evidences only indicate the gravitational interaction 
of dark matter. The detection of dark matter is yet a mystery. The only property of dark matter 
known to us is its relic abundance and is given by $\Omega_{\rm DM}h^2=0.119$~\cite{PLANCK}. Within the 
framework of the standard model (SM) it is impossible to accomodate a particle candidate of dark 
matter. 

Another aspect of SM is that it does not explain the non-zero neutrino mass which is confirmed by the 
phenomenon of neutrino oscillation observed in solar, atmospheric and reactor experiments \cite{neutosc}. 
This indicates towards new physics operative beyond electroweak scale since the SM of particle 
physics can not explain the same without introducing any extra ingredients. It is found that 
seesaw mechanisms are the most elegant scenarios for generating small neutrino masses. One of them 
is the type-I seesaw (or commonly refered as canonical seesaw \cite{typeI}) which relies on the 
existence of right-handed (RH) neutrinos. Despite the sucess of seesaw mechanisms in explaining light 
neutrino masses, they suffer from direct testability at LHC or at any other near future accelerator 
experiments. On the other hand, one can lower the scale of new physics by generating neutrino masses 
through radiative mechanisms~\cite{radiative_neu_mass}. 

The recently observed $3.5$ keV X-ray line signal in the spectrum of 73 galaxy 
clusters as reported by the XMM-Newton X-ray observatory~\cite{DMstab-bound,XMM-Newton} will provide 
a strong hint for non-gravitational Dark Matter interaction if confirmed by others. 
A few attempts have been already taken in order to explain this excess of X-ray signal 
by (i) decaying dark matter or annihilating dark matter, 
\cite{DM-decay-a,DM-decay-b,DM-decay-c,DM-decay-d,DM-decay-e,DM-decay-f,DM-decay-g, 
DM-decay-h,DM-decay-i,DM-decay-j,DM-decay-k,DM-decay-l,DM-decay-m}, 
(ii) magnetic dipolar dark matter (via upscattering of dark matter into an excited states and subsequent decay of 
excited states into dark matter and an X-ray photon),  
\cite{DM-dipole-a,DM-dipole-b,DM-dipole-c,DM-dipole-d,DM-dipole-e,DM-dipole-f,DM-dipole-g}, 
(iii) others  
\cite{DM-axion-a,DM-axion-b,DM-axion-c,DM-axion-d,DM-axion-e,DM-axion-f,DM-axion-g,
DM-axinos-a,DM-axinos-b,DM-axinos-c, 
DM-susy-a,DM-susy-b,DM-susy-c, 
DM-moduli-a,DM-moduli-b,
DM-others-a,DM-others-b,DM-others-c,DM-others-d,DM-others-e,DM-others-f,DM-others-g,DM-others-h,DM-others-i,DM-others-j}. 
From the recent explored scenarios it is found 
that such a photon signal, if the claimed excess persists, can be explained by late decay 
and/or annihilation of multi-keV mass dark matter, or decay of a second lightest metastable 
particle to lightest stable particle with a keV mass splitting. Such a monochromatic photon 
signal cannot be explained within the frame work of standard model (SM) of particle physics.

In this paper, we attempt to explain simultaneously the 3.5 keV X-ray line and non-zero light neutrino 
masses in a minimal extension of the SM while keeping the scale of new physics at the TeV scales. We 
extend the SM with three right-handed neutrinos ($N_1, N_2, N_3$), one singlet charged scalar $H^+$ and 
an extra Higgs doublet $\Sigma$. We also impose a discrete symmetry $Z_2 \times Z^\prime_2$ which does not 
allow to mix the RH-neutrinos. A small mixing between $N_1$ and $N_2$ is obtained by breaking the discrete 
symmetry softly. As a result a small mass splitting between $N_1$ and $N_2$ is created. This allows the 
next-to-lightest stable particle $N_2$ to decay to lightest stable particle $N_1$ through electromagnetic 
dipole moment operators. If the mass splitting between $N_1$ and $N_2$ is about 3.5 keV, then the emittted 
photon can be identified with the observed X-ray line by XMM-Newton X-ray observatory. Since the lepton 
number is violated by the Majorana mass of heavy RH neutrinos, the light neutrinos acquire their masses 
either at tree level or at loop level.         
 
The paper is organised as follows. In section-II and section-III we discuss in details a model of dipolar dark 
matter which not only give neutrino mass but also explain the 3.5 keV X-ray line. In section-IV, we discuss the 
relic abundance of dark matter. Constraints from direct detection of dark matter is given in section-V. Section-VI 
concludes.

\section{The Model for dipolar dark matter}
We augment the SM by introducing three right-handed fermions $N_{iR}$ which are 
essentially singlets under $SU(2)_L$. We also add a singlet charged scalar ($H^+$) and 
a Higgs doublet ($\Sigma $) to achieve the proposed objective. The masses of all these new 
particles are assumed to be of ${\cal O}(\rm TeV)$. A $~Z_2\times Z'_2~$ symmetry is also 
imposed in order to keep the lightest of right handed neutrinos stable, which also serves 
as a candidate of dark matter. The entire particle content, along with the 
quantum number assignments, is displayed in Table \ref{tab:SM}.
\begin{table}[!h]
\begin{center}
\caption{Particle content of the proposed Model.}
\label{tab:SM}
\begin{tabular}{|c|c|c|c|}
\hline
 &Field & $ SU(3)_C\times SU(2)_L\times U(1)_Y $ & $~Z_2\times Z'_2~$ \\
\hline
\hline
Fermions&$Q_L \equiv(u, d)^T_L$        & $(3, 2, 1/6)$         & + + \\
       &$u_R$                          & $(3, 1, 2/3)$       & + ~+  \\
       &$d_R$                          & $(3, 1, -1/3)$      & + ~+   \\
       &$\ell_L \equiv(\nu,~e)^T_L$    & $(1, 2, -1/2)$        & + ~+     \\
       &$e_R$                          & $(1, 1, -1)$          & + ~+    \\
       & $N_{1R}$                       & $(1, 1, 0)$           & -~ +       \\
       & $N_{2R}$                      & $(1, 1, 0)$           & +~ - \\
       & $N_{3R}$                      &$(1, 1, 0)$           & +~ + \\
\hline
Scalars&$\Phi$                         & $(1, 2, +1/2)$   & + ~+\\
       &$\Sigma$                       & $(1, 2, +1/2)$   & + ~-  \\
       &$H^+$                          & $(1, 1, +1)$     & - ~+   \\
\hline
\hline
\end{tabular}
\end{center}
\end{table}
We can then write the Lagrangian as:
 $${\cal L}={\cal L}_{\rm SM}+{\cal L}_{\rm New} $$
where ${\cal L}_{\rm SM}$ is the SM Lagrangian while the new physics Lagrangian ${\cal L}_{\rm New}$ contains
all terms containing any of the new particles including the right handed neutrinos and is given by:
\begin{eqnarray}
{\cal L}_{\rm New} & \ni & (Y_H)_{1\alpha}~ N_{1R}^T C \ell_{\alpha R} H^{+}  +
(Y_\Sigma)_{\alpha 2} \overline{\ell_{\alpha L}}  \tilde{\Sigma} N_{2R} \nonumber\\
&+& (Y_\nu)_{\alpha 3} \overline{\ell_{\alpha L}} \tilde{\Phi} N_{3R} + (Y_e)_{\alpha \beta} 
\overline{\ell_{\alpha L}} \Phi \ell_{\beta R} \nonumber\\
&+&\frac{1}{2} \overline{(N_{iR})^C} M_{Ni} N_{iR} + {\rm h.c.}+V(\Phi, \Sigma, H^+)
\end{eqnarray}
where the scalar potential can be given by:
\begin{eqnarray}
V(\Phi, \Sigma, H^+) &=& -\mu_\Phi^2|\Phi|^2 + M_\Sigma^2|\Sigma|^2 + M_{H}^2 |H^\pm|^2 \nonumber \\
&+&\lambda_\phi |\Phi|^4 + \lambda_{\Sigma} |\Sigma|^4 +\lambda_H |H^\pm|^4 \nonumber \\
&+&\lambda_{\Phi H} (\Phi^{\dag}\Phi)\, |H^\pm |^2 + \lambda_{\Sigma H} (\Sigma^{\dag}\Sigma) |H^\pm|^2 \nonumber \\
&+&f \, |\Phi|^2 |\Sigma|^2 + \frac{\lambda_{\Phi \Sigma}}{2} \left[ (\Phi^{\dag}\Sigma)^2  + h.c. \right] \,.  
\end{eqnarray}
Note that the $\Sigma$ is ascribed a positive mass-squared so that it does not acquire any vacuum expectation value (vev). 
Furthermore, we assume that $M_\Sigma > M_2$ and $M_{H^\pm} > M_1$ so that the decays $N_1 \to \ell_R H^+ $ and $N_2 \to 
\nu_L + \Sigma^0$ are kinematically forbidden. As a result $N_1$ and $N_2$ are individually stable. However, under $Z_2 
\times Z_2'$ symmetry, $N_3$ goes to itself and hence it is not stable. It can decay through the process: $N_3 \to \nu_i 
\Phi^0$. After Electro-Weak phase transition, the neutrino can acquire a small Majorana mass through the type-I seesaw 
anchored by $N_3$.

Notice that $Z_2 \times Z_2'$ symmetry doesn't allow $N_1$ and $N_2$ to mix with each other. We generate a small mixing 
between them by breaking $Z_2 \times Z_2'$ softly with: 
\begin{equation}
{\cal L}_{\rm soft} = \left[\mu_s\, \Sigma \Phi \,(H^+)^\ast + h.c.\right]\,,
\end{equation}
where $\mu_s$ has to be determined from the observed phenomenon. In particular, the breaking of $Z_2 \times Z_2'$ symmetry 
allows a mixing between $N_1$ and $N_2$. As a result $N_2$ can decay via $N_2 \to N_1 + \gamma$. The emitted photon 
can be identified with the recent observation of 3.5 keV X-ray line signal by the XMM-Newton X-ray observatory. 
  
\subsection{Constraints on new particles}
Since $\Sigma$ is a scalar doublet under $SU(2)_L$, it couples to $Z$-boson and hence can modify the decay width 
of $Z$-boson. Therefore, we take the masses of $\Sigma$ particles to be larger than $M_Z/2$. That means
\begin{equation}
M_\Sigma > 45 {\rm GeV}\,.
\end{equation}
The mass of the singlet chaged scalar $H^{\pm}$ is lower bounded by LEP. It is given to be~\cite{pdg} 
\begin{equation}
M_H > 80 {\rm GeV}\,.
\end{equation}

\subsection{Mass splitting between $N_1$ and $N_2$}
\begin{figure}[h]
 \centering
 \includegraphics[scale=0.56]{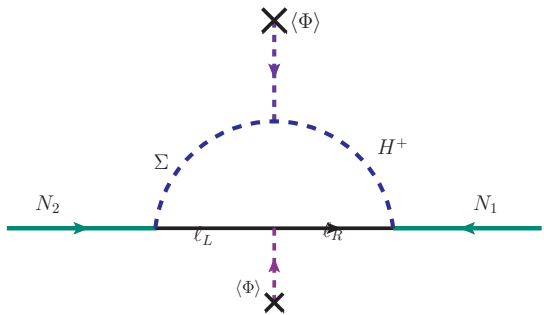}
 \caption{Mass splitting between $N_1$ and $N_2$ due to radiative correction through breaking of $Z_2 \times Z_2'$ symmetry.}
 \label{fig:2}
\end{figure}
As mentioned before, the breaking of $Z_2 \times Z_2'$ symmetry generates a mixing between $N_1$ and $N_2$. 
The mixing can be calculated from Fig.~\ref{fig:2} as: 
\begin{equation}\label{mass-splitting}
{\Delta M}_{12} = \frac{{Y^*_H}_{1\alpha}\, {Y_\Sigma}_{\alpha 2}}{16 \pi^2} 
                  \frac{\mu_s\, v_{\rm ew}\, m_\ell}{(M_\Sigma^2 - M_H^2)} \ln \left(\frac{M_\Sigma^2}{M_H^2}\right)
\end{equation}
As a result the Majorana mass matrix in the basis of $(N_1, N_2, N_3)$ can be given by:
\begin{equation}
\begin{pmatrix} M_1  & {\Delta M}_{12} & 0\cr\\
{\Delta M}_{12}  & M_2  & 0\cr\\
0 & 0 & M_3\end{pmatrix}
\end{equation}
Diagonalizing the mass matrix, we get the mass eigenvalues $M_1 + {\Delta M}_{12}$ and $ M_2-{\Delta M}_{12}$ 
and $M_3$. Thus the mass splitting between the two eigenstates $N_1$ and $N_2$ is given by 
\begin{equation}
\delta = 2 {\Delta M}_{12}\,.
\end{equation}
If we assume that $N_2$ is heavier than $N_1$, then the former can decay to latter by emitting a monochromatic 
photon with energy equal to the mass difference between them. Moreover, the life time of $N_2$ should be larger than the 
age of the universe so that it can decay in the current epoch. If the mass difference between $N_2$ and $N_1$ is 
of $\sim$ keV, then the recent observation of X-ray line can be accommodated in this beyond SM 
scenario.

\section{Magnetic dark matter (MDM) and monochromatic photon}
\begin{figure}[htb]
 \centering
 \includegraphics[scale=0.8]{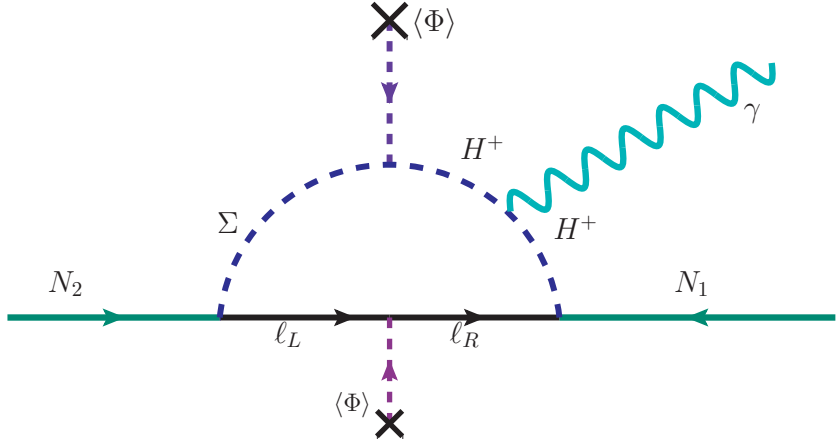}
 \caption{Diagram contributing to transient magnetic moment of right-handed neutrino in the proposed model.}
 \label{fig:1}
\end{figure}

Let us first discuss the electromagnetic properties of heavy RH neutrinos $N_{i}$ with photon before 
deducing the potential implications of the magnetic dipole moment operator in radiative decay $N_2 \to N_1 
+ \gamma$. Due to the Majorana nature, the diagonal magnetic moment of heavy Majorana neutrinos is zero. 
There is only transition magnetic moment for them. The electromagnetic coupling for the heavy neutrinos 
with photon, via dimension-five effective magnetic dipole moment (MDM) operator is
\begin{equation}
 \mathcal{L}_\text{\tiny MDM} =-\frac{i}{2}
  N_{R\,j}\,C^{-1}\, \mu_{jk} \sigma_{\alpha\beta}\, N_{R\,k} \,F^{\alpha\beta}
  +\text{h.c.} \quad (j\neq k).
\label{eqn:emdm}
\end{equation}
where ${\cal F}^{\alpha \beta}$ is the electromagnetic field tensor and $\mu_{12}$ is the transition magnetic 
moment between the first and second generation of heavy right-handed Majorana neutrinos $N_2$, $N_1$ which 
can be calculated from Fig. (\ref{fig:1}) as
\begin{equation}\label{magnetic-moment}
\mu_{12}=- \frac{-e}{64\,\pi^2} \frac{(Y_{H}^*\, Y_{\Sigma} \,\mu_s\,v_{\rm ew})}{M^2_\Sigma-M^2_H} 
\frac{m_\ell}{M^2_\Sigma}~{\cal I}_{\rm tot} 
\end{equation}
where 
\begin{eqnarray}
{\cal I}_{\rm tot} &\simeq& \int_{0}^{1} dx 
\bigg[\frac{x(1-x)^2}{(1-x) M^2_H/ M^2_\Sigma+x(x-1)M^2_2/ M^2_\Sigma+x} \nonumber \\
& &-\frac{x(1-x)^2}{(1-x) +x(x-1) M^2_2/ M^2_\Sigma+x M^2_H/ M^2_\Sigma}\bigg]\,.
\end{eqnarray}
Assuming $M_2 < M_{\Sigma}, M_H$ and with a typical set of values: $M_2 \simeq 100$ GeV, $M_\Sigma \simeq M_{H} = 384 {\rm GeV}$, 
the loop factor is estimated to be $4.4\times 10^{-5}$.

Since we assume the mass splitting between $N_1$ and $N_2$ is order of ${\rm keV}$, the decay of 
$N_2$ to $N_1$ can give rise a monochromatic $X$-ray line. The decay rate of $N_2\to N_1\gamma$ is estimated as:
\begin{eqnarray}
\Gamma(N_2\to N_1\gamma) &=& \frac{M_2^3 }{8 \pi}\left(1-\frac{M_1^2}{M_2^2 } \right)^3 |\mu_{12}|^2\nonumber\\
&=& \frac{ |\mu_{12}|^2}{\pi}\delta^3 \,\,\, ,
\label{ndecay}
\end{eqnarray}
where 
\begin{equation}
\delta \equiv  E_{\gamma} = \frac{ M_2} {2} \left(1-\frac{M_1^2}{M_2^2 } \right) 
\end{equation}
is the energy of the emitted photon, which is nothing but the mass difference between $N_1$ and $N_2$. For the 
observed $X$-ray line we use $\delta \sim 3.5 {\rm keV}$ and $\Gamma(N_2\to N_1\gamma) = 0.36 - 3.3 \times 10^{-52} 
{\rm GeV} (M_2/3.5 {\rm keV})$ for decaying dark matter~\cite{DMstab-bound,DM-dipole-g,DM-decay-f}. This implies that for $M_2 
= 100$ GeV, the decay rate is ${\cal O} (10^{-44}) {\rm GeV}$. In other words the life time of $N_2$ is 
${\cal O}(10^{19}) {\rm sec}$, which is larger than the age of the Universe. From Eqn. (\ref{ndecay}) one can 
estimate the required magnetic moment is $\mu_{12}={\cal O} (10^{-14})\, {\rm GeV}^{-1}$.

\subsection{Scale of new physics and collider search}
In order to fix the scale of new physics let us define the ratio: 
\begin{equation}
R \equiv \frac{\mu_{12} }{\delta}=\frac{e}{8} \frac{1}{M_\Sigma^2} \frac{I_{\rm tot}}{{\rm ln}(M_\Sigma^2/M_H^2)} \,. 
\end{equation}\label{ratio-eqn}
In the above equation $I_{\rm tot}$ can be evaluated numerically. In particular, for a 350 GeV DM mass, using 
$\delta=3.5 {\rm keV}$ and $\mu_{12}=2.46\times 10^{-14}\, {\rm GeV}^{-1}$ we get $M_\Sigma \approx M_H = 380 {\rm GeV}$. 
Thus the mass scale of the new particles are not far from the electroweak scale and hence can be searched at the collider.  
In particular, the charged scalars $H^\pm$ and $\Sigma^\pm$ are important. These particles can be pair produced at LHC 
via the exchange of SM Higgs particle. For example, $pp \to h \to H^+H^- \to e^+ e^- + {\rm missing~~ energy}$. Similarly, 
$pp \to h \to \Sigma^+\Sigma^- \to e^+ e^- + {\rm missing~~ energy}$. $\Sigma^\pm$ particles can also be detected through 
other decay processes, such as: $\Sigma^\pm \to W^\pm \Sigma^0 ({\Sigma^0}^*)\to f\bar{f} f_1 \bar{f_2}$, where $f,f_1,f_2$ 
are SM fermions.   

\subsection{Light neutrino Mass}
\begin{figure}[h]
 \centering
 \includegraphics[scale=0.56]{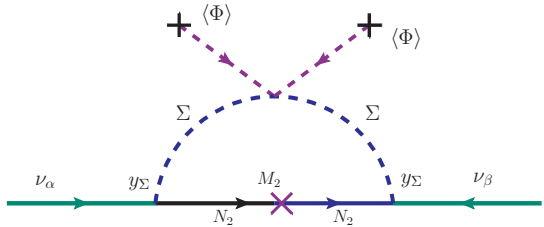}
 \caption{One loop diagram contributing to Majorana mass of light neutrinos.}
 \label{fig:nu}
\end{figure}

Through the electroweak (EW) phase transition, one of the Yukawa term $ (Y_\nu)_{\alpha 3} \overline{\ell_{\alpha L}} \tilde{\Phi} N_{3R}$ 
generates a Dirac mass term: $(M_D)_{\alpha 3}= (Y_\nu)_{\alpha 3} v_{\rm ew}$, where $v_{\rm ew} = \langle \Phi \rangle$ 
and $\alpha=e, \mu, \tau$. But the $N_3$ has a Majorana mass term $M_3 N_3 N_3$ which breaks lepton number by two units. 
As a result we get a Majorana mass matrix for light neutrinos to be: 
\begin{equation}
(m_\nu)_{\alpha \beta} =\frac{v^2}{M_3}  (Y_\nu)_{\alpha 3} (Y_\nu)_{\beta 3} 
\end{equation} 
Diagonalizing the above mass matrix we get the eigenvalues: ${\rm Tr}\left[ (Y_\nu)_{\alpha 3} (Y_\nu)_{\beta 3} \right] 
 v_{\rm ew}^2/M_3$, $0$, $0$. Thus at the tree level only one of the neutrinos is massive, say $\nu_3$. This is because of 
the exact $Z_2 \times Z_2'$ symmetry in the Lagrangian which prevents the Yukawa terms $\overline{\ell_\alpha} \Phi N_{1R}$ and 
$\overline{\ell_\alpha} \Phi N_{2R}$. On the other hand, the fact that $M_\Sigma^2 > 0$ prevents a vacuum expectation value 
for $\Sigma$ field, we can not generate a Majorana mass of light neutrinos at the tree level through the coupling: 
${Y_\Sigma}_{\alpha 2} \overline{\ell_{\alpha L}}  \tilde{\Sigma} N_{2R}$. However, the latter can be generated through one 
loop radiative correction diagram as shown in Fig.\ref{fig:nu}.
Because of the quartic coupling: $\frac{\lambda_{\Phi \Sigma}}{2} \left[ (\Phi^{\dag}\Sigma)^2  + h.c. \right]$, the 
EW phase transition generates a mass square splitting: $\lambda_{\Phi \Sigma} v^2$ between the real ($\Sigma^0_r$) 
and imaginary ($\Sigma^0_I$) components of $\Sigma^0$, the neutral component of $\Sigma$ field. As a result, the analytical expression 
for one loop-generated light neutrino mass is given by~\cite{ma_paper}:
\begin{eqnarray}
(m_\nu^{\rm loop})_{\alpha \beta}  &=&  \frac{(Y_\Sigma)_{\alpha 2} (Y_\Sigma)_{\beta 2}\, M_2}{16 \pi^2}
\bigg[\frac{M_{\Sigma_R}^2}{\left(M_{\Sigma_R}^2-M_2^2 \right)} \ln \left(\frac{M_{\Sigma_R}^2}{M_2^2} \right) \nonumber \\
& &\hspace{1.0cm}- \frac{M_{\Sigma_I}^2}{\left(M_{\Sigma_I}^2-M_2^2 \right)} \ln \left(\frac{M_{\Sigma_I}^2}{M_2^2} \right)  \bigg]
\end{eqnarray}
Diagonalising the above radiative mass matrix we get only one of the state massive, say $\nu_2$. Since the 
origin of the masses of the two eigen states $\nu_2$ and $\nu_3$ are different they can easily satisfy the 
solar and atmospheric mass splitting constraint.

\section{Relic abundance of $N_1$ and $N_2$}
The lightest stable particle (LSP), which is odd under $Z_2\times Z_2'$ symmetry, is the $N_1$ and hence 
behaves as a candidate of dark matter. The next to lightest stable particle (NLSP) is the $N_2$ and 
is stable on the cosmological time scale. It decays to $N_1$ through the electromagnetic coupling as shown in 
Fig. \ref{fig:1}. Since the observed $X$-ray line implies that the effective electromagnetic coupling of $N_1$ 
and $N_2$ with photon is extremely small, the life time of $N_2$ is larger than the age of the Universe. Therefore, 
$N_2$ is also a candidate of dark matter. When $N_2$ decays its density gets converted to $N_1$, while the 
net DM abundance remain intact. 

Since the electromagnetic coupling between $N_1$, $N_2$ and $\gamma$ is small, it can not bring the former two 
to thermal equilibrium. Note that to bring $N_1$ and $N_2$ to thermal equilibrium through electromagnetic coupling 
one needs the magnetic dipole moment $\mu_{12} \sim 10^{-4} {\rm GeV}^{-1}$~\cite{Masso:2009mu}. Hence the only way 
the right handed neutrinos are brought to thermal equilibrium is the interaction of $N_1$ with the singlet charged 
Higgs ($H^+$) and right-handed charged leptons ($\ell_R$) and that of $N_2$ with the $SU(2)_L$ doublets $\ell_L$ and 
$\Sigma$. For estimating the relic abundance of DM, the keV scale mass splitting between $N_1$ and $N_2$ is irrelevant. 
So for all practical purpose we assume $M_{N_1}=M_{N_2}=M_{\rm DM}$. As the temperature falls below the mass of DM, it 
decouples from the thermal bath. The decoupling temperature is given by $T_f=M_{\rm DM}/x_f$, where $x_f\approx 25$. The 
relevant cross-sections are given in appendix-B. The relic abundance obtained through co-annihilation of $N_1$ with $H^+$ 
and that of $N_2$ with $\Sigma$ can be given by~\cite{Griest&seckel_prd91}
\begin{equation}
\Omega_{\rm DM}h^2 = \frac{1.09\times 10^9 {\rm GeV}^{-1}}{g_*^{1/2}M_{\rm Pl}} \times \frac{1}{J(x_f)}
\end{equation}  
where 
\begin{equation}
J(x_f)= \int_{x_f}^\infty \frac{ \langle \sigma|v|\rangle_{\rm eff} }{x^2}
\end{equation}
with
\begin{equation}
\langle \sigma|v|\rangle_{\rm eff}=\langle \sigma|v|\rangle_{\rm eff}^{N_1} + \langle \sigma|v|\rangle_{\rm eff}^{N_2}\,.
\end{equation}
In the above equation the effective co-annihilation cross-sections of $N_1$ with $H^{+}$ and that of $N_2$ with that of 
$\Sigma$ are given as:
\begin{equation}
\langle \sigma|v|\rangle_{eff}^{N_i}=\langle \sigma|v|\rangle^{\small N_i-X} g_i g_X (1+\Delta_X)^{3/2}
\times \left( \frac{e^{-x\Delta_X}}{g_{\rm eff}^2} \right) 
\end{equation}
where $g_i, i=1,2$ represents the internal degrees of freedom of $N_1$ and $N_2$, $g_X$ represents the internal degrees 
of freedom of $X$-particle ($H^+, \Sigma$) co-annihilating with $N_i$, $i=1,2$ and 
\begin{equation}
g_{\rm eff}=g_i+g_X(1+\Delta_X)^{3/2}e^{-x\Delta_X}\,.
\end{equation}
where $\Delta_X=(M_X-M_{\rm DM})/M_{\rm DM}$.

\begin{figure}[htb]
 \centering
 \includegraphics[scale=0.56]{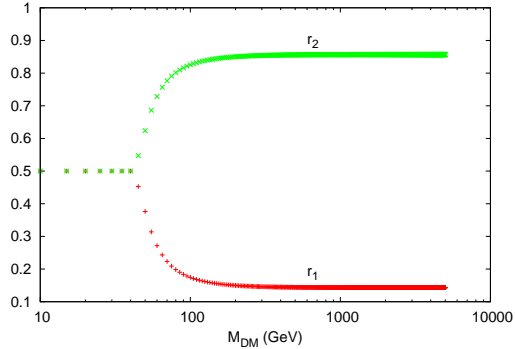}
 \caption{$r_1$ and $r_2$ as a function of $M_{\rm DM}$ for $Y_\Sigma=Y_H$. For simplicity we set $M_{\rm DM}=M_{X}$.}
 \label{fig:branching_ratio}
\end{figure}

\begin{figure}[h]
 \centering
 \includegraphics[scale=0.56]{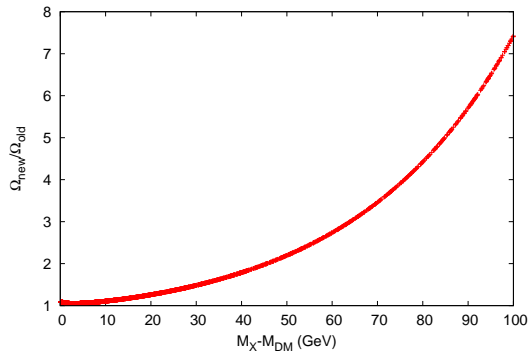}
 \caption{Ratio of dark matter abundance as a function of $M_{\rm X}-M_{\rm DM}$ for a typical mass 
of DM, $M_{\rm DM}=1000 {\rm GeV}$ and $Y_\Sigma=Y_H =0.55$ }
 \label{fig:omega_ratio}
\end{figure}

\begin{figure}[h]
 \centering
 \includegraphics[scale=0.56]{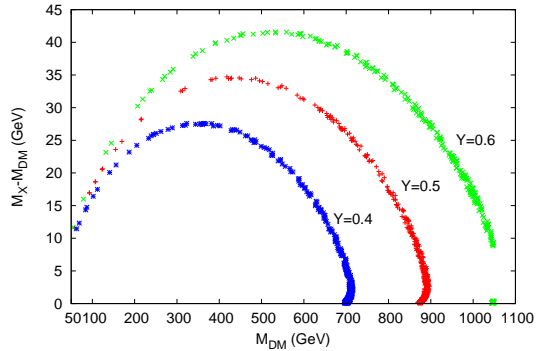}
 \caption{Observed relic abundance in the plane of $M_{\rm X}-M_{\rm DM}$ versus $M_{\rm DM}$ for different values of the 
couplings, i.e., $Y=0.4,0.5,0.6$.}
 \label{fig:mass_splitting_DM}
\end{figure}

To estimate the contribution of $N_1$ and $N_2$ to relic abundance we define the ratios: 
\begin{eqnarray}
r_1=\frac{ \langle \sigma|v| \rangle_{N_1} } {\langle \sigma|v| \rangle_{N_1}+ \langle \sigma|v| \rangle_{N_2}}\nonumber\\
r_2=\frac{ \langle \sigma|v| \rangle_{N_2} } {\langle \sigma|v| \rangle_{N_1}+ \langle \sigma|v| \rangle_{N_2}}
\end{eqnarray}
We have shown $r_1$ and $r_2$ as a function of $M_{\rm DM}$ in Fig. (\ref{fig:branching_ratio}) for $Y_\Sigma=Y_H = Y$. 
Moreover, for simplicity we assume the masses of $H^+$ and $\Sigma$ to be $M_{\rm X}$.
We define $\Omega_{\rm old} h^2$ as the relic abundance of DM assuming $M_{\rm DM} = M_{\rm X}$. Then we check the 
effect of mass splitting on the relic abundance. This is shown in Fig. (\ref{fig:omega_ratio}). From Fig. (\ref{fig:omega_ratio}) 
we see that relic abundance increases with the rise of mass splitting between the DM and X-particles. This is due to the 
fact that relic abundnace is inversly propertional to the effective cross-section, which suffers a Boltzmann suppression, i.e., 
$\langle \sigma|v| \rangle \propto e^{-x\Delta_X}$. Therefore, the effective cross-section decreases with the increase of mass 
splitting.         
Next we show the dependency of the observed relic abundance simultaneously on the mass of DM and the mass splitting between 
DM and X-particles for various couplings. This is demonstrated in Fig. (\ref{fig:mass_splitting_DM}) for $Y=0.4,0.5,0.6$. As 
expected for small mass splitting the co-annihilation is large between $N_1$, $N_2$ and the corresponding X-particles. Therefore 
we get large number of points near $M_{\rm X}-M_{\rm DM} \approx 0$. On the other hand, for a given mass of DM, if the coupling 
increases then we need relatively large mass splitting so that the effective cross-section decreases to get the right relic abundance. 
This feature can be easily read from Fig. (\ref{fig:mass_splitting_DM}) as we move from $Y=0.4$ to $Y=0.6$. For light mass of DM, say 
$M_{\rm DM} < 100 {\rm GeV}$, the cross-section increases with the further decrease of DM mass. Therefore, we need a large mass splitting 
so that the effective cross-section decreases to give the correct relic abunance.  

\section{Direct detection of dark matter and constraints}
\begin{figure}[h]
 \centering
 \includegraphics[scale=0.56]{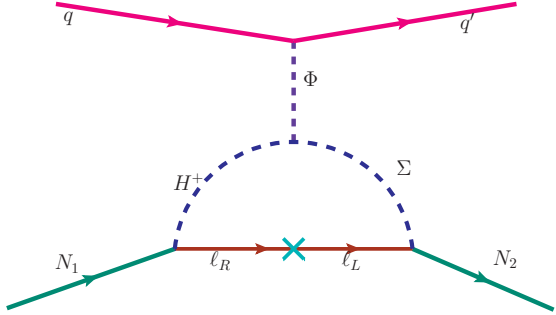}
 \caption{Elastic scattering of magnetic dipolar dark matter with target nuclei.}
 \label{fig:dirdet_signal}
\end{figure}

In this section, we wish to constrain the model parameters from the direct detection experiments for 
dark matter and LUX experiment \cite{lux} is one such experiment considered in our present analysis. 
Since the mass splitting between $N_1$ and $N_2$ is only 3.5 keV, so for all practical purposes we consider 
the spin independent DM-nucleon interaction $N_1 n \to N_2 n$, mediated by SM Higgs exchange, to be elastic. 
The Feynman diagram for DM-nucleon interaction for direct ditection is shown in Fig.\,\ref{fig:dirdet_signal}. 
We denote the effective coupling between $N_2-N_1-\Phi$ entering into this interaction as $\lambda_{\rm eff}$ 
and is given by:
\begin{eqnarray}
\lambda_{\rm eff}\simeq \frac{- \mu_S}{16 \pi^2} Y_\Sigma Y^*_H m_\ell \mathcal{F}\left(M^2_H, M^2_\Sigma, m^2_\ell \right) \, ,
\end{eqnarray}
where 
\begin{eqnarray}
&&\mathcal{F}\left(M^2_H, M^2_\Sigma, m^2_\ell \right) =\frac{1}{M^2_H-M^2_\Sigma} \times \nonumber \\
&&\bigg[\frac{1}{\frac{m^2_\ell}{M^2_H}-1}\ln\left(\frac{m^2_\ell}{M^2_H} \right) 
-\frac{1}{\frac{m^2_\ell}{M^2_\Sigma}-1}\ln\left(\frac{m^2_\ell}{M^2_\Sigma} \right) \bigg]\,.
\end{eqnarray}
Thus the spin independent DM-nucleon cross-section can be given as:
\begin{equation}
\sigma_{SI}= \frac{\mu_n^2}{4\pi} \lambda_{\rm eff}^2 \left(\frac{f_n m_n}{v_{\rm ew} } \right)^2 \frac{1}{M_h^4}
\end{equation}
where $\mu_n = M_{N_1}m_n/(M_{N_1}+ m_n)$ is the reduced DM-nucleon mass with $m_n=0.946 GeV$ and $M_h=125 {\rm GeV}$ is 
the SM Higgs mass. The effective coupling between the SM Higgs and nucleon is given by: $f_n m_n/v_{\rm ew}$ which depend 
upon the quark content of the nucleon for each quark flavour. The Higgs-nucleon coupling $f_n$ is given by: 
\begin{equation}
f_n= \sum_q f_q = \frac{m_q}{m_n} \langle n | \bar{q} q | n \rangle
\end{equation}
where the sum is over all quark flavors. In the present analysis, we have used $f_n=0.32$ \cite{phi-nucl-coupl} 
though its value can lie within a range $f_n=0.26-0.33$ \cite{phi-nucl-coupl-range}. Using $M_\Sigma \approx M_H=100 {\rm GeV}$ 
and $Y_H = Y_\Sigma =0.5 $ we get $\sigma_{\rm SI} \approx {\cal O} (10^{-59}) {\rm cm^2}$, which is much smaller than the 
current stringent limit: $\sigma_{\rm SI}=7.5 \times 10^{-46}\,\mbox{cm}^2$ from LUX.

\section{Conclusion}
In this paper we have shown that the observed $3.5$ keV $X$-ray line signal and neutrino mass can be 
explained simultaneously in a minimal extension of the SM. We extend the SM with three right-handed 
neutrinos ($N_1, N_2 N_3$), a charged scalar $H^+$ and a Higgs doublet ($\Sigma$) which transforms 
non-trivially under a discrete symmetry $Z_2 \times Z^\prime_2$. The lightest odd particle is $N_1$ 
and behaves as a candidate of DM. The next to lightest odd particle is $N_2$ whose mass is separated 
from $N_1$ by 3.5 keV. As a result the total DM of the universe is a mixture of $N_1$ and $N_2$ abundance. 
However, the discrete symmetry is broken softly so that $N_2$ decays to $N_1+ \gamma$ in the present 
universe, converting its abundance to $N_1$ while keeping the total DM abundance intact. The $\gamma$ can 
be identified with the observed 3.5 keV $X$-ray line. Since $N_3$ couples to the SM lepton and Higgs doublet, 
a small Majorana mass for one of the light neutrinos is generated through the canonical seesaw mechanism. 
The other neutrinos get their masses at one loop level.  
    
We show that the required relic abundance of DM implies the masses of new particles $H^\pm$ and $\Sigma^\pm$ 
are not far from the electroweak scale. So these partciles can be pair produced and their subsequent decay 
can be searched at LHC. 

We noticed that the direct detection cross-section of DM with the nucleon is highly suppressed. Therefore, the 
constraints on the model parameters are almost negligible.

\section{Acknowledgments}
The work of Sudhanwa Patra is partially supported by the Department of Science and Technology, 
Govt. of India under the financial grant SB/S2/HEP-011/2013. Narendra Sahu is partially 
supported by the Department of Science and Technology, Govt. of India under the financial 
Grant SR/FTP/PS-209/2011. 

\section*{\large Appendix A: Calculation for transitional magnetic moment for \hspace*{4cm} magnetic dark matter 
                 decay $N_2 \to N_1 \gamma$}
\begin{figure}[h]
\centering
\includegraphics[scale=0.6]{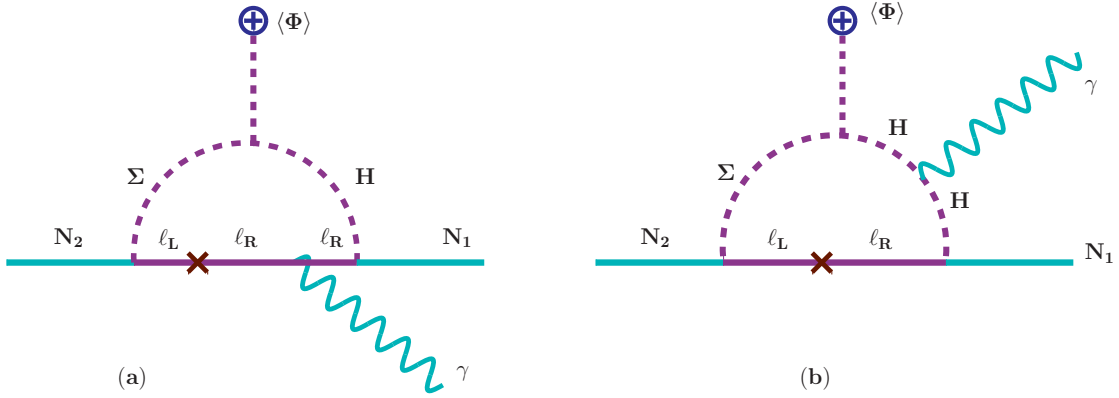}
\caption{Feynman diagram for magnetic dark matter decay: $N_2 \to N_1 \gamma$.}
\label{feyn:magmmom-app}
\end{figure}

The effective Lagrangian accounting the interaction between photon and two neutral fermions states 
is given as
\begin{equation}
 \mathcal{L}_\text{\tiny MDM} =-\frac{i}{2}
  N_{R\,j}\,C^{-1}\, \mu_{jk} \sigma_{\alpha\beta}\, N_{R\,k} \,F^{\alpha\beta}
  +\text{h.c.} \quad (j\neq k).
\end{equation}
We present here the Feynman calculation for transitional magnetic moment between two neutral fermion states 
$N_2$ and $N_1$ leading to the decay $N_2 \to N_1 \gamma$ explaining the $3.5$ keV X-ray line signal. 
The possible Feynman diagrams which yield transitional magnetic moment ${\bf \mu}_{12}$ and decay process 
$N_2 \to N_1 \gamma$ is shown in Fig.\,\ref{feyn:magmmom-app}. 
Let us assign the four-vector momenta of $N_2$, $N_1$ and proton by $p_1$, $p_2$ and $k$, respectively.
The Feynman amplitude for the magnetic dark matter decay shown in Fig.\,\ref{feyn:magmmom-app}(a) can 
be written as
\begin{eqnarray}
{\cal M}^{(a)}&=&\int \frac{d^4 k}{(2\pi)^4} \overline{u(p_2)} \left(i\,Y_{H}^*\, P_R\right)\, 
S_{\ell}(k-q) \left(-i\, e\,\gamma_\mu \epsilon^\mu(q)\right)\,S_{\ell}(k) (i\, Y_{\Sigma}\, P_R) \,u^c(p_1) \nonumber \\
&&\hspace*{5.5cm} \times \Delta_{H}(p_1-k)\,\, \left(\mu_s\,v_{\rm ew}\right)\, \, \Delta_{\Sigma}(p_1-k)   \nonumber \\
&=& \overline{u(p_2)} \left(i\,\Gamma^{(a)}_\mu \right) \epsilon^\mu (q) u^c(p_1) \, ,
\end{eqnarray}
where $i\,\Gamma^{(a)}_\mu$ is factored out to be
\begin{eqnarray}
i\,\Gamma^{(a)}_\mu &=&\left(-e\,Y_{H}^*\, Y_{\Sigma}\,\mu_s\,v_{\rm ew}\right) \times \nonumber \\
&&\hspace*{0.2cm}\int \frac{d^4 k}{(2\pi)^4} 
\frac{P_R (k\hspace{-0.19cm}/-q\hspace{-0.18cm}/ +m_\ell)\, \gamma_\mu (k\hspace{-0.2cm}/ +m_\ell)\, P_R}
{[\{(k-q)^2-m^2_\ell\} \{k^2-m^2_\ell\} \{(p_1-k)^2-M^2_H)\} \{(p_1-k)^2-M^2_\Sigma)\}]} \nonumber \\
&=&\left(-e\,Y_{H}^*\, Y_{\Sigma} \,\mu_s\,v_{\rm ew}\right) \int \frac{d^4 k}{(2\pi)^4} 
\mathcal{D}^{-1} \cdot \mathcal{N}_\mu  
\end{eqnarray}
We denote $\mathcal{D}^{-1}$ and $\mathcal{N}_\mu$ as follows
\begin{eqnarray}
&&\mathcal{D}^{-1}= \frac{1}{[\{(k-q)^2-m^2_\ell\} \{k^2-m^2_\ell\} \{(p_1-k)^2-M^2_H)\} \{(p_1-k)^2-M^2_\Sigma)\}]} \\
&&\mathcal{N}_\mu=P_R (k\hspace{-0.19cm}/-q\hspace{-0.18cm}/ +m_\ell)\, \gamma_\mu (k\hspace{-0.2cm}/ +m_\ell)\, P_R\,.
\end{eqnarray}

After doing some simpler algebra, we get
\begin{eqnarray}
\label{eqn:gamma2}
\Gamma^{(a)}_{\mu}&=&\frac{\left(-e\,Y_{H}^*\, Y_{\Sigma} \,\mu_s\,v_{\rm ew}\right)}{M^2_\Sigma-M^2_H} \times 
\int \frac{d^4 k}{(2\pi)^4} \, \bigg[ \frac{N_\mu}{\{(k-q)^2-m^2_\ell\}\{k^2-m^2_\ell\} 
\{(p_1-k)^2-M^2_\Sigma\}}\nonumber \\
&-&\frac{\mathcal{N}_\mu}{\{(k-q)^2-m^2_\ell\}\{k^2-m^2_\ell\} 
\{(p_1-k)^2-M^2_H\}}  \bigg] 
\nonumber \\
&=&\frac{\left(-e\,Y_{H}^*\, Y_{\Sigma} \,\mu_s\,v_{\rm ew}\right)}{M^2_\Sigma-M^2_H}[I_1-I_2]
\end{eqnarray}
The integral now becomes

\begin{eqnarray*}
I_1&=&\frac{m_\ell}{16\,\pi^2} \int_{0}^{1} dx \int_{0}^{1-x} dy 
~S^{-2}_1[2(q\,x+p_1\,y)_\mu-q\hspace{-0.2cm}/ \gamma_\mu]P_R \nonumber \\
&=&\frac{1}{16\,\pi^2} \frac{m_\ell}{M^2_\Sigma}\int_{0}^{1} dx \int_{0}^{1-x} dy 
~\Omega_1\,[2\,y\,p_{1\,\mu}-2\,p_{1\,\mu}]P_R \nonumber \\
&=&\frac{1}{8\,\pi^2} \frac{m_\ell}{M^2_\Sigma} p_{1\,\mu} 
\int_{0}^{1} dx \int_{0}^{1-x} dy ~\Omega_1\,[y-1]P_R
\end{eqnarray*}
Here $S^{-2}_1=M^{-2}_\Sigma\, \Omega_1$, $\Omega_1=[Z_N \left(y-y^2 -xy\right) 
     -(1-y)Z_\ell-y]^{-1}$.
Since the term proportional to $p_{2\,\mu}-p_{1\,\mu}$ gives vanishing 
contribution and writing $p_{1\,\mu}=\frac{1}{2}[p_{2\,\mu}+p_{1\,\mu}+q_\mu]$, 
the integral becomes
\begin{eqnarray}
I_1=\frac{1}{16\,\pi^2} \frac{m_\ell}{M^2_\Sigma} (p_1+p_2)_\mu 
\int_{0}^{1} dx \int_{0}^{1-x} dy ~\Omega_1\,[y-1]P_R
\end{eqnarray}
Similarly, one can express the second integral ${\cal I}_2$ by replacing $M_\Sigma$ by $M_H$. 

Now we can write the loop factor as
\begin{eqnarray}
\label{eqn:gamma3}
i\,\Gamma^{(a)}_{\mu}&=& \frac{\left(-e\,Y_{H}^*\, Y_{\Sigma} \,\mu_s\,v_{\rm ew}\right)}{M^2_\Sigma-M^2_H} 
\bigg[ \frac{1}{16\,\pi^2} \frac{m_\ell}{M^2_\Sigma} (p_1+p_2)_\mu 
\int_{0}^{1} dx \int_{0}^{1-x} dy ~\Omega_1\,(y-1)P_R  \nonumber \\
&-& \frac{1}{16\,\pi^2} \frac{m_\ell}{M^2_\Sigma} (p_1+p_2)_\mu 
\int_{0}^{1} dx \int_{0}^{1-x} dy ~\Omega_2\,(y-1)P_R \bigg]  \nonumber \\
&=& \frac{-e}{16\,\pi^2} \frac{(Y_{H}^*\, Y_{\Sigma} \,\mu_s\,v_{\rm ew})}{M^2_\Sigma-M^2_H} 
\frac{m_\ell}{M^2_\Sigma}~{\cal I} ~(p_1+p_2)_\mu\, P_R \nonumber \\
&=&{\cal A}_{12}^{({\bf a})}~(p_1+p_2)_\mu
\end{eqnarray}
where the integral ${\cal I} $ is
$${\cal I}=\int_{0}^{1} dx \int_{0}^{1-x} dy\,(y-1) [\Omega_1-\Omega_2]$$ and
$${\cal A}_{12}^{({\bf a})}= \frac{-e}{16\,\pi^2} \frac{(Y_{H}^*\, Y_{\Sigma} \,\mu_s\,v_{\rm ew})}{M^2_\Sigma-M^2_H} 
\frac{m_\ell}{M^2_\Sigma}~{\cal I}$$
From Gordan Identity, we can write
\begin{eqnarray}
{\cal A}_{12}^{({\bf a})}~\overline{u(p_2)}\, (p_1+p_2)_\mu u(p_1)={\cal A}_{12}^{({\bf a})}~ \overline{u(p_2)} \left[\,2 M_N\,\gamma_\mu 
-i\, \sigma_{\mu\,\nu}(p_2-p_1)^\nu \right] u(p_1) 
\end{eqnarray}
The analytical expression for transitional magnetic moment between two nearly degenerate heavy RH neutrinos, as derived 
from the effective operator ${\cal A}_{12}^{({\bf a})}\overline{N_{1}} \sigma^{\mu \nu} N_{2} F_{\mu \nu}$ following Feynman diagram 
Fig.\,\ref{feyn:magmmom-app}(a), as 
\begin{equation}
{\cal A}_{12}^{({\bf a})}= \frac{-e}{16\,\pi^2} \frac{(Y_{H}^*\, Y_{\Sigma} \,\mu_s\,v_{\rm ew})}{M^2_\Sigma-M^2_H} 
\frac{m_\ell}{M^2_\Sigma}~{\cal I} 
\end{equation}
Following the same method of Feynman calculation, one can derive the relevant contributions of transition magnetic moment 
depicted in Fig.\,\ref{feyn:magmmom-app}(b). At the end, the analytical expression for transitional magnetic moment including 
relevant Feynman diagrams can be put in the following expression
\begin{eqnarray}
{\cal \mu}_{12}&=& {\cal A}_{12}^{({\bf a})}+{\cal A}_{12}^{({\bf b})} \nonumber \\
&\simeq &\frac{-e}{16\,\pi^2} \frac{(Y_{H}^*\, Y_{\Sigma} \,\mu_s\,v_{\rm ew})}{M^2_\Sigma-M^2_H} 
\frac{m_\ell}{M^2_\Sigma}~{\cal I}_{\rm tot}\, .
\end{eqnarray}
\section*{\large Appendix B:\, Cross-section calculation for relic abundance of DM}
We present here the cross-section calculation involving magnetic dark matter and possible 
coannihilation channels relevant for relic abundance estimation. The relic abundance of $N_1$ 
is obtained through its co-annihilation with $H^+$ via the following processes: 
\begin{eqnarray}
&&\langle \sigma|v \rangle (N _{1R}+H^+\Leftrightarrow \ell_R+\phi) 
       = \frac{(Y_H)_{1\alpha}^2}{64\pi M_H(s-m_\alpha^2)^2}\frac{m_f^2}
{\langle\phi\rangle^2} s \sqrt{s} \left(1-\frac{M_\phi^2}{s}\right)^2 \\
&&
\langle \sigma|v \rangle (N_{1R}+H^+\Leftrightarrow \ell_R+Z ) = \frac{1}{32\pi M_H}\left(\frac{g(Y_H)_{1\alpha}(T_3-sin^2\theta_W\, Q)}
{cos\theta_W (s-m_\alpha^2+i\epsilon)}\right)^2s\sqrt{s}\left(1-\frac{m_Z^2}{s}\right)^2 \\
&&
\langle \sigma|v \rangle (N_{1R}+H^+\Leftrightarrow \ell_R+\gamma)=\frac{e^2}{32\pi M_H}\frac{(Y_H)_{1\alpha}^2}{(s-m_\alpha^2)^2}s\sqrt{s}
\end{eqnarray} 
where $\alpha = e, \mu, \tau$
 
The relic abundance of $N_2$ is obtained through its co-annihilation with $\Sigma$ particles via the following processes are
\begin{eqnarray}
&&\langle \sigma|v \rangle (N_{2R}+\Sigma^0\Leftrightarrow l_L+W) =  \frac{1}{64\pi M_\Sigma} 
         \frac{g^2(Y_\Sigma)_{2\alpha}^2}{\sqrt{s}} \left(1-\frac{m_W^2}{s}\right)^2  \\
&&\langle \sigma|v \rangle (N_{2R}+\Sigma^0\Leftrightarrow \nu_L+Z) =\frac{1}{128\pi M_\Sigma}
         \frac{g^2(Y_\Sigma)_{2\alpha}^2}{cos^2\theta_W\sqrt{s}}\left(1-\frac{m_Z^2}{s}\right)^2  \\
&&\langle \sigma|v \rangle (N_{2R}+\Sigma^-\Leftrightarrow \nu_L+W) =\frac{1}{64\pi M_\Sigma}
          \frac{g^2(Y_\Sigma)_{2\alpha}^2}{(s-m_\alpha^2+i\epsilon)^2}s\sqrt{s}\left(1-\frac{m_W^2}{s}\right)^2 \\
&&\langle \sigma|v \rangle (N_{2R}+\Sigma^-\Leftrightarrow e_L+Z) =\frac{1}{32\pi M_\Sigma}
         \frac{g^2(Y_\Sigma)_{2\alpha}^2(T_3-sin^2\theta_WQ)^2}{cos^2\theta_W(s-m_\alpha^2+i\epsilon)^2}s\sqrt{s}\left(1-\frac{m_Z^2}{s}\right)^2 \\
&&\langle \sigma|v \rangle (N_{2R}+\Sigma^-\Leftrightarrow e_L+\gamma) =\frac{e^2(Y_\Sigma)_{2\alpha}^2}{32\pi M_\Sigma 
          (s-m_\alpha^2+i\epsilon)^2}s\sqrt{s}   \\
&&
\langle \sigma|v \rangle (N_{2R}+\Sigma^-\Leftrightarrow e_R+\phi) =\frac{(Y_\Sigma)_{2\alpha}^2}{64\pi M_\Sigma(s-m_\alpha^2+i\epsilon)^2}
         \frac{m_f^2}{<\phi>^2}s\sqrt{s}\left(1-\frac{M_\phi^2}{s}\right)^2
\end{eqnarray}

\end{document}